\documentclass[prl,twocolumn,superscriptaddress,showpacs,amssymb,aps,floatfix]{revtex4}
\newcommand{\be}{\begin{equation}}
\newcommand{\ee}{\end{equation}}
\newcommand{\bea}{\begin{eqnarray}}
\newcommand{\eea}{\end{eqnarray}}
\usepackage{graphicx}
\usepackage{float}
\usepackage{amsmath}
\usepackage{color}
\usepackage{dcolumn}
\usepackage{bm}
\usepackage{hyperref}
\usepackage{braket}
\usepackage{enumerate}

\begin{document}
\title{Time-of-Flight Measurements as a Possible Method to Observe Anyonic Statistics}

\author{R. O. Umucal\i lar}
\affiliation{%
Deparment of Physics, Mimar Sinan Fine Arts University, 34380 Sisli, Istanbul, Turkey
}%
\author{E. Macaluso}
\affiliation{%
INO-CNR BEC Center and Dipartimento di Fisica, Universit$\grave{a}$ di Trento, 38123 Povo, Italy
}%
\author{T. Comparin}
\affiliation{%
INO-CNR BEC Center and Dipartimento di Fisica, Universit$\grave{a}$ di Trento, 38123 Povo, Italy
}%
\author{I. Carusotto}%
\affiliation{%
INO-CNR BEC Center and Dipartimento di Fisica, Universit$\grave{a}$ di Trento, 38123 Povo, Italy
}%

\date{\today}

\begin{abstract}
We propose a standard time-of-flight experiment as a method for observing the anyonic statistics of quasiholes in a fractional quantum Hall state of ultracold atoms. The quasihole states can be stably prepared by pinning the quasiholes with localized potentials and a measurement of the mean square radius of the freely expanding cloud, which is related to the average total angular momentum  of the initial state, offers direct signatures of the statistical phase. Our proposed method is validated by Monte Carlo calculations for $\nu=1/2$ and $1/3$ fractional quantum Hall liquids containing a realistic number of particles. Extensions to quantum Hall liquids of light and to non-Abelian anyons are briefly discussed.
\end{abstract}


\maketitle
{\it Introduction.---}
The usual exchange statistics, which classifies particles into bosons and fermions, is enriched in two dimensions (2D). In 2D, the many-particle wave function can, in principle, acquire an arbitrary statistical phase factor $\exp(i\phi_{\rm st})$ upon particle exchange, which can be different from the usual $\pm 1$ factor defining bosons and fermions \cite{Wu,Leinaas}.  Particles having this unusual fractional exchange statistics are called anyons \cite{anyon review}. In the presence of topologically degenerate ground states, the phase factor  when anyons are braided around each other can even be replaced by non-Abelian transformations acting on the ground state manifold \cite{Wilczek 1}, with interesting potential applications in topological quantum computing \cite{anyon computation review}.

Among the 2D systems where anyons appear naturally, fractional quantum Hall (FQH) systems are, perhaps, the most commonly studied ones \cite{anyon review,Laughlin paper,Yoshioka}. Quasihole and quasiparticle excitations of an FQH system are known to exhibit anyonic character \cite{Wilczek 2}. Although the FQH effect was originally observed in 2D electron gases under a magnetic field \cite{FQH experiment}, analogue systems where interacting neutral particles experience synthetic magnetic fields \cite{synthetic field atoms,topological photonics} are emerging as promising platforms for studying the FQH physics. Ultracold atomic \cite{cold atoms} and photonic systems \cite{light review}, being prime examples of such analogue systems, are advantageous over the electronic ones in that they offer a highly controllable environment. In these systems, it might be possible to pin and braid anyons using localized potentials for particles \cite{Zoller anyon,Lewenstein FQH,manybody braiding phases,Mueller}. 

While the fractional charge of (Abelian) anyons in FQH systems has been experimentally observed via shot-noise measurements \cite{fractional charge exp}, no clear-cut evidence of the exchange statistics is yet available. Although interferometric measurements performed, so far, in electronic systems are highly suggestive of anyonic statistics \cite{anyon exp}, they still lack a unique interpretation \cite{anyon exp interpretation}. More recent studies on the interferometry of Abelian anyons include a more detailed modeling of the usual Fabry-P\'erot setups, which accounts for competing effects~\cite{Halperin Fabry-Perot} and the proposition of Hanbury Brown-Twiss interferometry to probe anyon correlations~\cite{Campagnano}. Building on earlier proposals \cite{non-Abelian proposals}, experiments pointing at non-Abelian properties were also performed~\cite{Willett}. Interferometric schemes for detecting the statistical phase were also developed for ultracold atomic~\cite{Zoller anyon} and photonic~\cite{Mueller} systems. Recently, as a slightly different approach, proposals for detecting Haldane's fractional exclusion statistics~\cite{Haldane exclusion}, which is intimately connected to the braiding statistics, have appeared in the solid state~\cite{Papic exclusion}, ultracold atomic~\cite{Cooper exclusion}, and photonic contexts~\cite{onur2017}.

In this Letter, we propose a much simpler time-of-flight (TOF) measurement \cite{cold atoms} as a way to observe the statistical phase of an FQH liquid of ultracold atoms initially prepared in a quasihole state with Abelian braiding statistics. The suggested experimental procedure involves creating and pinning the quasiholes with localized potentials and suddenly releasing the atomic cloud to measure the density distribution after time of flight for one- and two-quasihole states. As the average total angular momentum of the initial state can be mathematically related both to the TOF mean square radius \cite{Ho Mueller} and to the Berry phase \cite{Berry phase,Wilczek 2} associated with quasihole braiding, a measurement of the former provides information on the latter quantity. As a key advantage over previous interferometric proposals \cite{Zoller anyon,Mueller}, ours does not require physically moving quasiholes and is based on a standard TOF measurement on a \emph{static} system.

Different, yet related aspects of the anyonic character of quasiholes have been addressed in recent works: the fractionalization of angular momentum has been discussed in \cite{Jain} for test particles immersed in an ultracold atomic FQH system and, very recently, in \cite{Yakaboylu} for impurities interacting with a bosonic bath. Signatures of anyonic statistics in the correlation functions of an expanding gas of anyons have been suggested in \cite{Buljan}.

{\it Model System.---}
We consider a generic FQH system with interacting neutral particles in a synthetic magnetic field $B$, which is uniform and perpendicular to the 2D plane of motion. Such a system, with $N$ particles of mass $M$, can be described by the Hamiltonian
\begin{multline}
H_{\rm FQH} = \sum_{i=1}^N \frac{(-i\hbar\nabla_i-\mathbf{A})^2}{2M}+g_{\rm int}\sum_{i<j}\delta^{(2)}(\mathbf{r}_i-\mathbf{r}_j),
\label{H_FQH}
\end{multline}
where $\mathbf{A}(\mathbf{r}) = B\hat{\mathbf{z}}\times\mathbf{r}/2$ is the synthetically created symmetric gauge vector potential. The strength of repulsive contact interactions is given by $g_{\rm int}>0$.

The eigenstates of the noninteracting Hamiltonian are the Landau levels separated by the cyclotron energy $\Delta E = \hbar B/M$. When the typical interaction energy $g_{\rm int}/l_{\mathrm{B}}^2$, with $l_{\mathrm{B}} = \sqrt{\hbar/B}$ the magnetic length, is sufficiently smaller than $\Delta E$, it is reasonable to make the approximation that only the lowest Landau level (LLL) is occupied. The wave function of a single-particle eigenstate in the LLL with angular momentum $n\hbar$ is $\psi_n(\zeta) = \zeta^{n}e^{-|\zeta|^{2}/4}/(\sqrt{2\pi 2^n n!}l_{\mathrm{B}})$, where $\zeta = (x + iy)/l_{\mathrm{B}}$ is the complex-valued coordinate of the particle.

For the many-particle system, one can define the filling fraction $\nu = N/N_{\Phi}$ as the ratio between the number of particles $N$ and the number of magnetic flux quanta $N_{\Phi}$, which corresponds to the filling of the Landau levels in the noninteracting case. For a fractional filling $\nu=1/m$, the exact nondegenerate ground state of the interacting Hamiltonian $H_{\rm FQH}$ at a total angular momentum $\mathcal{L}_z = mN(N-1)\hbar/2$ is described by the Laughlin wave function \cite{Laughlin paper,Zoller anyon,Bosonic FQH}
\bea
\Psi_{\rm FQH}(\zeta_1, \ldots, \zeta_N) \propto \prod_{j<k}(\zeta_j-\zeta_k)^m e^{-\sum_{i = 1}^N|\zeta_i|^2/4},\label{WF_FQH}
\eea
where $\zeta_i$ is the complex-valued coordinate of the $i$th particle. For bosons (fermions) $m$ must be even (odd) for the symmetry of the wave function to be correct. In what follows, we will focus on the two exemplary cases with $m = 2$ and $3$. The $m = 3$ wave function is the ansatz proposed by Laughlin to describe the FQH effect for electrons at filling $\nu = 1/3$~ \cite{Laughlin paper}. The $m = 2$ bosonic wave function appeared, instead, in the context of rotating ultracold atoms \cite{Zoller anyon,Bosonic FQH,Cooper review} and was theoretically found to be the absolute ground state in the presence of a uniform synthetic magnetic field and a weak trapping potential \cite{external potential,onur2017}. In a similar setup with fermionic atoms, the ground state will be the $m = 3$ Laughlin state.

The Laughlin wave function (\ref{WF_FQH}) has zero-energy excitations known as quasiholes which obey anyonic exchange statistics in the thermodynamic limit~\cite{Yoshioka,Wilczek 2}. When two quasiholes are exchanged, the many-body wave function acquires the phase $\phi_{\rm st} = \nu \pi$. Numerical studies show that quasiholes (qh) can be pinned by repulsive piercing potentials created with lasers in ultracold atomic systems \cite{Zoller anyon,Lewenstein FQH}. Such a potential term can be represented by a sum of delta potentials as $V_{\rm qh} = V_0\sum_{i=1}^{N_{\rm qh}}\sum_{j=1}^{N}\delta^{(2)}({\bf r}_j-{\bf R}_i)$, where $N_{\rm qh}$ is the total number of repulsive potentials and ${\bf R}_i$ is the position of the $i$th localized potential with strength $V_0$. 

According to exact diagonalization of small systems~\cite{SuppMat1}, the ground state of the total Hamiltonian $H_{\rm qh} = H_{\rm FQH}+V_{\rm qh}+V_{\rm trap}$ including suitable pinning and trapping \cite{external potential,onur2017} potentials is not affected by the details of the potentials and is well represented by the following one- and two-quasihole wave functions for $N_{\rm qh} = 1$ and 2, respectively,
\bea
\Psi_{\rm 1qh}(\{\zeta_i\},\mathcal{R}_1) \propto  \prod_{i=1}^N(\zeta_i\!-\!\mathcal{R}_1)\Psi_{\rm FQH}(\zeta_1, \ldots, \zeta_N) \label{1QH wf},\\
\Psi_{\rm 2qh}(\{\zeta_i\},\{\mathcal{R}_j\}) \propto  \prod_{i=1}^N\prod_{j =1}^2(\zeta_i\!-\!\mathcal{R}_j)\Psi_{\rm FQH}(\zeta_1, \ldots, \zeta_N),
\label{2QH wf}
\eea
where $\mathcal{R}_{1,2}$ are the complex positions of the quasiholes determined by the positions ${\bf R}_{1,2}$ of the localized potentials. 
From the experimental perspective, once the ultracold atomic cloud is prepared in the Laughlin ground state, which is assumed to be sufficiently separated from both the gapped bulk excitations and low-lying edge excitations so that thermal fluctuations do not spoil it, one can adiabatically prepare the quasihole states by slowly increasing the strength $V_0$ of the repulsive potentials and then slowly moving them to the desired position in space \cite{Zoller anyon}. Alternatively, one may also consider directly cooling down the gas in the presence of the potentials. 

{\it Braiding phase and total angular momentum.---}
In our system with quasiholes, the braiding phase corresponds to the difference between the Berry phases the many-body wave function acquires after a quasihole is moved along a closed path with or without another quasihole enclosed by the path \cite{Wilczek 2,Yoshioka}. Provided the quasiholes remain sufficiently far apart from each other, the braiding phase does not depend on the details of the path; therefore, we can consider a circular path of radius $R$, cyclically parametrized by the angular coordinate $\theta$. We further assume that the second quasihole (if present) is pinned at the origin. The Berry phase \cite{Berry phase}, in this case, becomes
\be
\varphi_{\rm B}(R) = i\oint_{R} \langle \Psi(\theta)|\partial_{\theta}|\Psi(\theta)\rangle d\theta,
\label{Berry phase}
\ee
where $|\Psi(\theta)\rangle$ refers to the one- or two-quasihole states (\ref{1QH wf})--(\ref{2QH wf}).

We now relate the Berry phase (\ref{Berry phase}) to the expected value of total angular momentum $\langle L_z \rangle$, by first writing the action of the partial derivative $\partial_\theta$ on the state $|\Psi(\theta)\rangle$ as $\partial_{\theta}|\Psi(\theta)\rangle = \lim_{\delta\theta\rightarrow 0}\{[|\Psi(\theta+\delta\theta)\rangle-|\Psi(\theta)\rangle]/\delta\theta\}$. Since rotating the quasihole by $\delta\theta$ is equivalent to rotating the whole many-body system by the same angle (modulo a $2\pi$-periodic phase factor, linearly dependent on $\delta\theta$), the state $|\Psi(\theta+\delta\theta)\rangle$ can be represented using the rotation generator $L_z$ as $|\Psi(\theta+\delta\theta)\rangle = \exp(-iL_z\delta\theta/\hbar)|\Psi(\theta)\rangle$. Expanding the rotation operator  for small $\delta\theta$ as $\exp(-iL_z\delta\theta/\hbar)\simeq 1-iL_z\delta\theta/\hbar$, we see that $\partial_{\theta}|\Psi(\theta)\rangle = -(iL_z/\hbar)|\Psi(\theta)\rangle$, which implies
\be
\varphi_{\rm B}(R) = \frac{1}{\hbar}\oint_{R} \langle \Psi(\theta)|L_z|\Psi(\theta)\rangle d\theta = \frac{2\pi}{\hbar}\langle L_z\rangle,
\label{Berry phase ang mom}
\ee
where the expectation value of $\langle L_z\rangle$ is taken with respect to a wave function having a quasihole with fixed radial coordinate $R$ but an arbitrary angular coordinate. 
This remarkable expression relates a quantity resulting from an adiabatic motion, that is, braiding, to a stationary property of a quantum mechanical state, that is, the average total angular momentum \cite{note on L_z}.

\begin{figure*}[ht]
\includegraphics[width=1.0\textwidth]{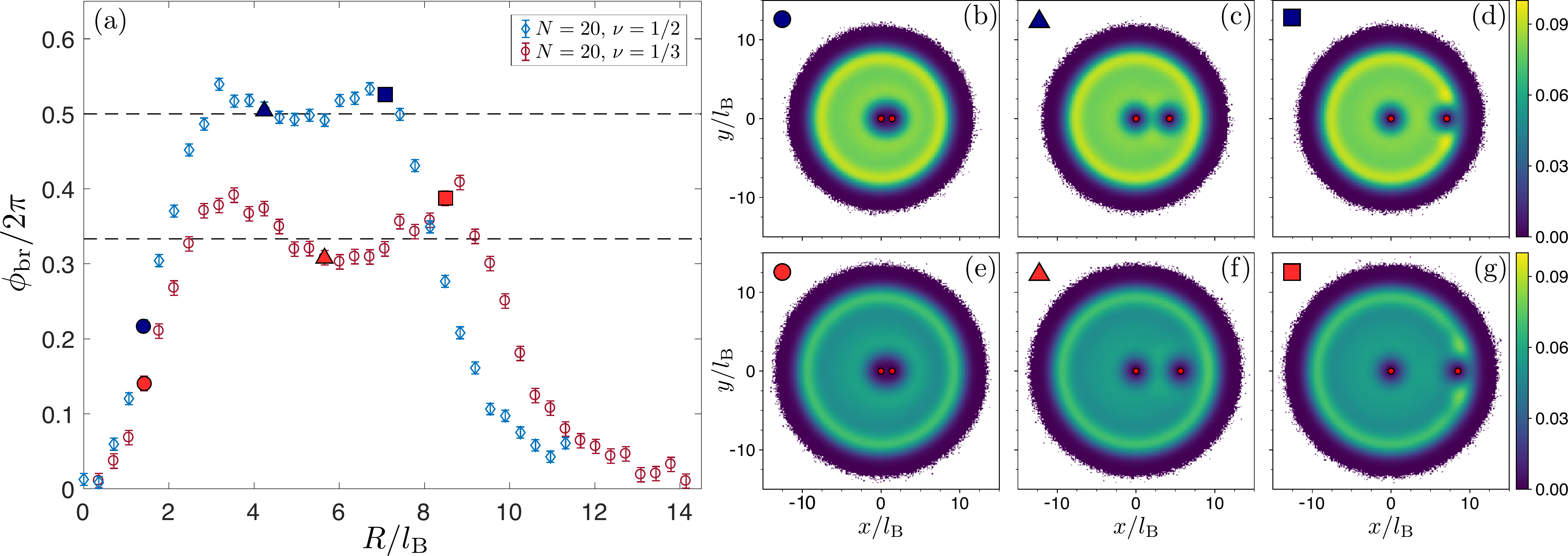}
\caption{(a) Quasihole braiding phase $\phi_{\rm br}$ as a function of the distance $R = |{\bf R}_{1}-{\bf R}_{2}|$ between two quasiholes for systems of $N=20$ particles at filling $\nu=1/2$ (blue diamonds) and $\nu=1/3$ (red circles), where one of the quasiholes is fixed in the origin ($|{\bf R}_{2}|=0$). Error bars represent statistical uncertainties on the data. Density profiles characterizing some two-quasihole states are given for $\nu=1/2$ in (b)--(d) and for $\nu=1/3$ in (e)--(g). The position of the outer quasihole is fixed along the $x$ axis at $x_{1}/\sqrt{2} l_{\mathrm{B}} = 1,3,5$ in (b)--(d) and at $x_{1}/\sqrt{2} l_{\mathrm{B}} = 1,4,6$ in (e)--(g).}
\label{fig:Phi_N20_m2m3}
\end{figure*}

Equation~(\ref{Berry phase ang mom}) also provides an experimental route for measuring the braiding (br) phase given by the Berry phase differences $\phi_{\rm br}(R) = \varphi_{\rm B}^{\rm 2qh}(R)-\varphi_{\rm B}^{\rm 1qh}(R)$ yielding
\be
\phi_{\rm br}(R) = \frac{2\pi}{\hbar}(\langle L_z\rangle^{\rm 2qh}-\langle L_z\rangle^{\rm 1qh}),
\label{Braiding phase ang mom}
\ee
where the superscripts 1qh and 2qh refer to one- and two-quasihole states, respectively. Equation (\ref{Braiding phase ang mom}) shows that the braiding phase can simply be determined by measuring the average total angular momentum for two quantum states and taking the difference, without any need to actually braid quasiholes. The fact that the braiding phase is defined in Eq.~(\ref{Braiding phase ang mom}) only up to an integral multiple of $2\pi$ does not preclude highlighting the fractional statistics.

As an alternative to the braiding phase, one may choose to directly measure the statistical (st) phase $\phi_{\rm st}(R) = \phi_{\rm br}(R)/2$ involving the adiabatic exchange of two quasiholes \cite{Lewenstein FQH,Mueller}. In our proposal, this would correspond to measuring the angular momentum in the two cases of (a) two quasiholes pinned at diametrically opposite positions each at a distance $R/2$ from the origin and (b) a single quasihole pinned at a radius $R/2$. In particular, one can write
\be
\phi_{\rm st}(R) = \frac{\pi}{\hbar}(\langle L_z\rangle_{{\rm op}}^{\rm 2qh}-2\langle L_z\rangle^{\rm 1qh}) + \pi m \frac{N(N-1)}{2},
\label{Statistical phase}
\ee
where $\langle L_z\rangle_{{\rm op}}^{\rm 2qh}$ is the average total angular momentum of the two-quasihole state with diametrically opposite (op) quasiholes and the last term compensates the phase factor picked up by the quasihole wave functions after a $\pi$ rotation~\cite{SuppMat2}. Although we will evaluate the statistical phase only for the case of oppositely located quasiholes, Eq.~\eqref{Statistical phase} can be generalized to configurations in which quasiholes are pinned at generic positions. However, such a generalization requires three different measurements of $\langle L_{z} \rangle$, instead of two~\cite{SuppMat2}.

{\it Time-of-flight measurement.---}
The average total angular momentum of a cloud of cold atoms occupying the LLL can be determined by just measuring the mean square radius $\langle r^2\rangle$ of the density distribution of atoms in the trap or, even easier, after a time-of-flight expansion for a duration $t$ once the pinning and trapping potentials and synthetic fields are suddenly turned off \cite{Ho Mueller,SuppMat3}
\begin{equation}
\langle r^2\rangle_{{\rm TOF}} = \frac{1}{N}\bigg(\frac{\hbar t}{\sqrt{2}Ml_{\mathrm{B}}} \bigg)^2\bigg(\frac{\langle L_z\rangle}{\hbar} + N\bigg)
= \bigg(\frac{\hbar t}{2Ml_{\mathrm{B}}^2} \bigg)^2 \langle r^2\rangle.
\label{r2tof Lz}
\end{equation}
Note that for this self-similar TOF expansion to be valid, the interactions between particles should be negligible during the expansion, but the initial state can well be a highly correlated one. This omission of interaction effects can be justified, for instance, whenever the system can be described within the LLL approximation \cite{LLL expansion}.

Combining Eq.~(\ref{Braiding phase ang mom}) with the relation displayed in Eq.~(\ref{r2tof Lz}) between the in-trap average total angular momentum $\langle L_z\rangle$ and $\langle r^2\rangle_{{\rm TOF}}$, we obtain the fundamental experimental observable yielding the braiding phase
\be
\phi_{\rm br}(R) \simeq 2\pi N\bigg(\frac{\sqrt{2}Ml_{\mathrm{B}}}{\hbar t}\bigg)^2(\langle r^2\rangle^{\rm 2qh}_{{\rm TOF}}-\langle r^2\rangle^{\rm 1qh}_{{\rm TOF}}),
\label{r2 braiding phase}
\ee
which is, again, defined up to an integral multiple of $2\pi$. Similarly, the corresponding observable for $\phi_{\rm st}(R)$ can be found by using Eqs. (\ref{Statistical phase}) and (\ref{r2tof Lz}).

{\it Numerical Results.---}
In this section, we substantiate our conclusions by presenting estimates for $\phi_{\rm br}(R)$ calculating the in-trap mean square radius $\langle r^2\rangle$, related to $\langle r^2\rangle_{\rm TOF}$ through Eq.~(\ref{r2tof Lz}). Our numerical calculations are based on the analytical wave functions (\ref{1QH wf})--(\ref{2QH wf}) and we use a Monte Carlo (MC) technique~\cite{code} to compute $\langle r^2\rangle$ and the density profile~\cite{SuppMat4}. As a further check, we performed exact diagonalization calculations for smaller $N$ so as to benchmark the MC results and verify that the ground state wave functions for suitable pinning and trapping potentials match the analytical wave functions~\cite{SuppMat1}.

\begin{figure}[htbp]
\includegraphics[width=0.45\textwidth]{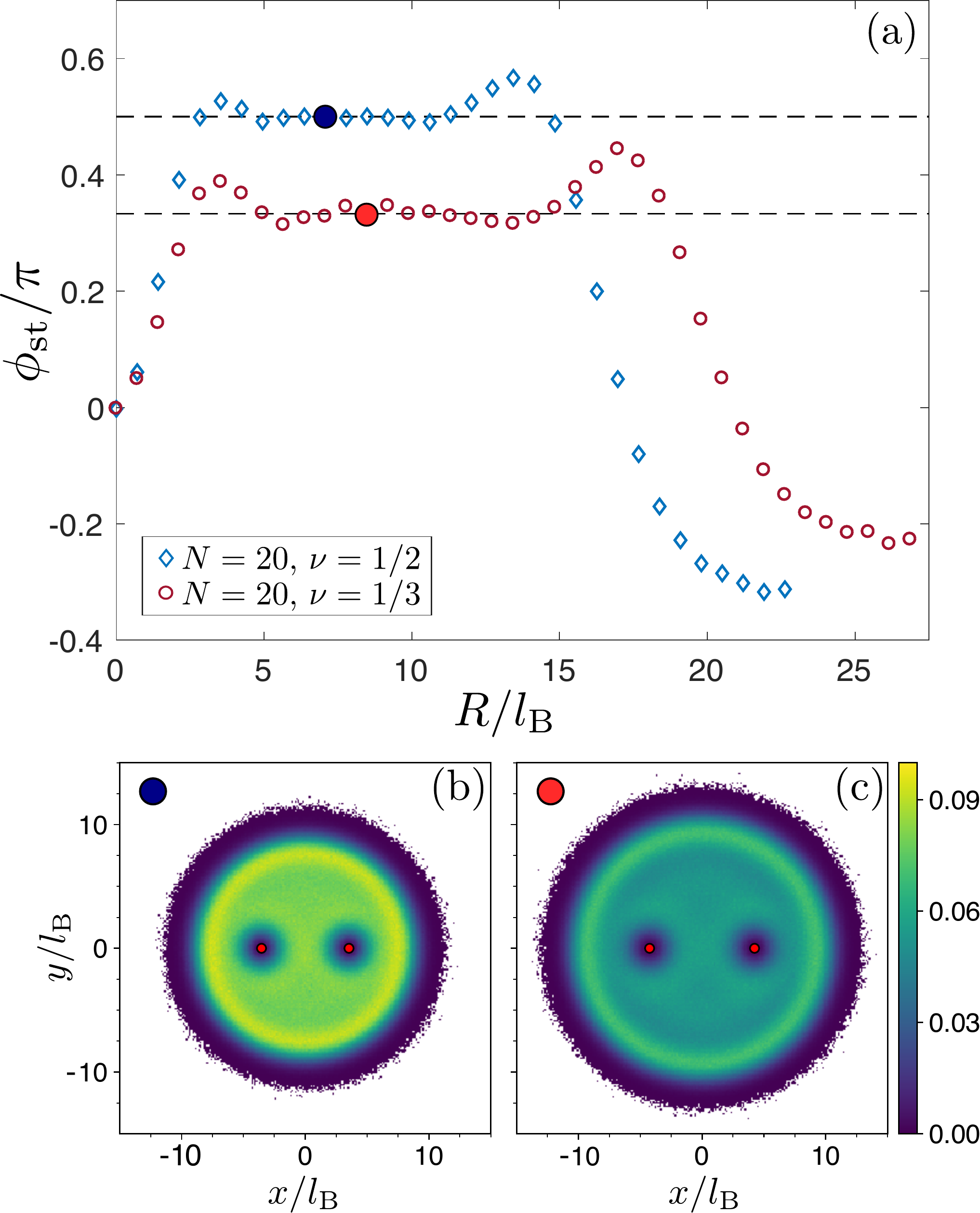}
\caption{(a) Statistical phase $\phi_{\rm st}$ as a function of the distance $R = |{\bf R}_{1}-{\bf R}_{2}|$ between two quasiholes for systems of $N=20$ particles at filling $\nu=1/2$ (blue diamonds) and $\nu=1/3$ (red circles), where the quasiholes are located at diametrically opposite positions ($x_{1} = -x_{2} = R/2$). Error bars are smaller than the symbol size. (b) Density profile for $\nu=1/2$ with quasiholes located along the $x$ axis at $x_{1} = -x_{2} = 2.5 \sqrt{2} l_{\mathrm{B}}$. (c) Density profile for $\nu=1/3$ with quasiholes located along the $x$ axis at $x_{1} = -x_{2} = 3 \sqrt{2} l_{\mathrm{B}}$.}
\label{fig:Phi_N20_m2m3_oppositeQHs}
\end{figure}

We consider two configurations of two-quasihole states, where  the distance between two quasiholes is denoted by $R = |{\bf R}_1-{\bf R}_2|$ in each case. In Fig. \ref{fig:Phi_N20_m2m3}, one of the quasiholes is located at the center, so we calculate $\phi_{\rm br}(R) = 2\pi N(\langle r^2\rangle^{\rm 2qh}-\langle r^2\rangle^{\rm 1qh})/(\sqrt{2}l_{\mathrm{B}})^2$ determined by Eqs.~(\ref{Braiding phase ang mom}) and (\ref{r2tof Lz}). In Fig. \ref{fig:Phi_N20_m2m3_oppositeQHs}, two quasiholes are located at diametrically opposite positions, so the relevant quantity is $\phi_{\rm st}(R) = \pi N(\langle r^2\rangle^{\rm 2qh}_{\rm op}-2\langle r^2\rangle^{\rm 1qh})/(\sqrt{2}l_{\mathrm{B}})^2+\pi N+\pi mN(N-1)/2$ from Eqs.~(\ref{Statistical phase}) and (\ref{r2tof Lz}). 

In the calculations for $N=20$-particle systems shown in Fig.~\ref{fig:Phi_N20_m2m3}(a), a clear plateau is seen for $\nu=1/2$ at the expected fractional value $\phi_{\rm br}/2\pi = 1/2$, accompanied by small bumps at its ends. These bumps are more pronounced in the $\nu = 1/3$ case, where the plateau is not fully visible, and can be related to perturbations in the FQH cloud density.

On one hand, at small $R$ [filled circles in Fig.~\ref{fig:Phi_N20_m2m3}(a)], the phase behavior reflects the density deformation induced by the quasihole in the origin shown in Figs.~\ref{fig:Phi_N20_m2m3}(b) and \ref{fig:Phi_N20_m2m3}(e). Such a deformation is not sensitive to the cloud size that increases with $N$ and directly reflects the size of the quasihole. Especially for $\nu=1/3$ the bump in $\phi_{\rm br}(R)$ precisely matches the position of the peak in the density profile~\cite{PRB33_2221,PRL110_186801, PRB89_115124}. On the other hand, the bump visible at large $R$ [filled squares in Fig.~\ref{fig:Phi_N20_m2m3}(a)] is related to the density increase in the vicinity of the cloud edge [Figs.~\ref{fig:Phi_N20_m2m3}(d) and \ref{fig:Phi_N20_m2m3}(g)]. The scaling of the bump position with $N$ and that of the bump visibility with $\nu$ confirm the behavior of the density maximum: the former scales as $\sqrt{N}$, while the latter increases with decreasing $\nu$.

While smaller clouds give qualitatively similar results albeit with quantitatively more pronounced deviations, these calculations prove that, for $\nu=1/2$, an $N=20$-particle system is already big enough to properly measure the anyonic statistics of quasiholes. On the other hand, for $\nu=1/3$, the bigger effective size of quasiholes requires larger systems to clearly observe the plateau in the braiding phase. Since larger particle numbers typically require a higher relative precision in measuring the angular momentum, a useful alternative option is to consider the second configuration with quasiholes at diametrically opposite positions~\cite{Lewenstein FQH,Mueller}: such a configuration allows us to maximize the quasihole distance by exploiting the full extension of the bulk region. In this way, it is possible to obtain a clear plateau in the statistical phase also for $\nu=1/3$ and $N=20$ particles, as displayed in Fig.~\ref{fig:Phi_N20_m2m3_oppositeQHs}(a).

{\it Conclusion.---}
In this Letter, we argued that a standard measurement of the static density profile in the trap or after time of flight is sufficient to observe the anyonic statistics of quasiholes in a gas of ultracold atoms in the FQH regime. We showed that the mean square radius of the cloud in the presence of one or two quasiholes is directly related to the braiding and statistical phases. Numerical calculations of the braiding phase $\phi_{\rm br}$ as a function of the distance between quasiholes for a reasonable number of particles ($N=20$) clearly display a plateau region, for which the quasiholes are sufficiently far away from each other and from the edge of the cloud. Except for small finite-size deviations, the value of the plateau is very close to the expected one $\phi_{\rm st} = \phi_{\rm br}/2 = \nu\pi$, giving a clear signature of the quasihole anyonic statistics.

A possible extension of our protocol to the case of non-Abelian anyons is also the subject of ongoing studies. The key difference is that the Berry phase is replaced by its Wilczek-Zee generalization~\cite{anyon computation review}, which depends on a matrix of inner products of the form $\langle \Psi_{\alpha} (\theta) | \partial_{\theta} | \Psi_{\beta} (\theta) \rangle$. Indices $\alpha$ and $\beta$ label the degenerate quasihole ground states peculiar to non-Abelian phases. In order to generalize our scheme, we will first need to verify the identification of $\partial_{\theta} |\Psi_{\alpha} (\theta) \rangle$ with $L_{z} |\Psi_{\alpha} (\theta) \rangle$ and then to connect the angular momentum matrix elements $\langle \Psi_{\alpha}(\theta)| L_z | \Psi_{\beta} (\theta) \rangle$ with certain real-space observables like $\langle r^{2} \rangle$ considered in the current Letter. Such a real-space approach might be appealing, particularly in view of the possibility of using a generalized plasma analogy~\cite{Nayak_generalized_plasma}. From the perspective of reproducing our proposal in the non-Abelian context, it looks promising to consider the $p_{x} + i p_{y}$ model of topological superconductors~\cite{anyon computation review}, as the Moore-Read state~\cite{MooreRead_Pfaffian}, representing the simplest FQH state with non-Abelian statistics, can be described through the $p$-wave pairing of composite fermions \cite{Moller}.

Further work will extend these results to FQH liquids of photons in cylindrical set-ups such as the twisted resonators of~\cite{jon simon}, for which the far-field intensity profile of the light emission provides the optical analog of time-of-flight imaging of ultracold atomic clouds. A first task will be to identify suitable schemes to generate stable quasiholes states, e.g., by generalizing the frequency-dependent incoherent pumping scheme of~\cite{onur2017} in the presence of pinning potentials piercing the cavity. We then expect that the braiding phase of quasiholes can again be extracted from the expectation value of the angular momentum.

\begin{acknowledgments}
This work was supported by the EU-FET Proactive grant AQuS, Project No. 640800, and by the Autonomous Province of Trento, partially through the project ``On silicon chip quantum optics for quantum computing and secure communications". Discussions with N. Cooper, L. Mazza, T. Ozawa, H. M. Price, and R. Santachiara are warmly acknowledged.
\end{acknowledgments}


\begin{thebibliography}{99}
\bibitem{note on L_z} This result was obtained in \cite{manybody braiding phases} for the special case when the many-body wave function can be written in the LLL. The same relation has appeared in a different system investigated in \cite{Yakaboylu}.
\end{thebibliography}
\end{document}


\title{Supplemental Material}
\maketitle

\section{\label{sec:phase_factor} Extracting the statistical phase}
With respect to the expression for the braiding phase, the one for the statistical phase is characterized by an additional term [cf. Eqs.~(7) and (8) in the main text]. Here we explain the origin of such a term.

In general, the one-quasihole and the two-quasihole wave functions read
\begin{equation}
\begin{split}
\Psi_{\rm 1qh}(\{\zeta_i\},\mathcal{R}_1) &\propto  \prod_{i=1}^N(\zeta_i\!-\!\mathcal{R}_1)\Psi_{\rm FQH}(\zeta_1, \ldots, \zeta_N) ,\\
\Psi_{\rm 2qh}(\{\zeta_i\},\{\mathcal{R}_j\}) &\propto  \prod_{i=1}^N\prod_{j =1}^2(\zeta_i\!-\!\mathcal{R}_j)\Psi_{\rm FQH}(\zeta_1, \ldots, \zeta_N) ,
\end{split}
\end{equation}
where $\{\zeta_i\}$ and $\{\mathcal{R}_j\}$ denote the particle and the quasihole coordinates in the complex plane, respectively, and $\Psi_{\rm FQH}(\zeta_1, \ldots, \zeta_N)$ is the Laughlin wave function,
\begin{equation}
\Psi_{\rm FQH}(\zeta_1, \ldots, \zeta_N) \propto \prod_{j<k}(\zeta_j-\zeta_k)^m e^{-\sum_{i = 1}^N|\zeta_i|^2/4}.
\label{WF_FQH}
\end{equation}

As a rotation by an angle $\theta$ maps the particle coordinates $\{\zeta_i\}$ into $\{\zeta_i e^{-i \theta}\}$, its effect on the Laughlin wave function is simply described by the appearance of a global phase factor
\begin{equation}
\phi_{L} (\theta) = m \frac{N(N-1)}{2} \theta.
\end{equation}
This clearly shows the $2 \pi$ periodicity of the Laughlin wave function. The one- and two-quasihole wave functions [Eqs.~(1) and (2)], in addition to the phase $\phi_{L} (\theta)$, acquire an extra phase $\phi_{n-\mathrm{qh}}(\theta) = nN\theta$, due to the polynomial pre-factors.

While all these phase factors become irrelevant for the evaluation of the braiding phase $\phi_{\mathrm{br}}$, which manifests itself after a $2 \pi$-rotation, one has to account for them when computing the statistical phase $\phi_{\mathrm{st}}$, and in particular when dealing with fermions. In this case $m$ is odd and therefore $\phi_{L} (\pi)$ is in general not an integer multiple of $2 \pi$.

Upon the exchange of the two quasiholes through a $\pi$-rotation, $\Psi_{\rm 2qh}(\{\zeta_i\},\{\mathcal{R}_j\})$ takes three different phase factors. In addition to the term $\phi_{L} (\pi) + \phi_{2-\mathrm{qh}}(\pi)$ mentioned above, the two-quasihole wave function also acquires the statistical phase $\phi_{\mathrm{st}}$ and a geometric phase accounting for the motion in the parameter space of the single quasiholes. Therefore one has to get rid of the other contributions to extract $\phi_{\mathrm{st}}$. Subtracting twice the Berry phase of the one-quasihole configuration allows us to remove both the geometric phase associated with the adiabatic motion of the single quasiholes and the contributions to the phase coming from the polynomial pre-factors. However, this introduces an extra $- \phi_{L} (\pi)$ term, which appears in Eq.~(8) in the main text.

\section{\label{sec:TOF} Relation between time-of-flight and in-trap averages}

Here we derive Eq.~(9) of the main text. In doing this, we first need to compute the Fourier transform of the lowest Landau level (LLL) wave function.

\subsection{Fourier transform of the LLL wave function}

The Fourier transform of $\psi_n({\bf r}) = (r/l_{\mathrm{B}})^n e^{i n\phi}e^{-r^2/4l^2_B}/(\sqrt{2\pi 2^n n!}l_{\mathrm{B}})$ is calculated as
\begin{multline}
\tilde{\psi}_n({\bf k}) = \frac{1}{2\pi}\int e^{-i{\bf k}\cdot{\bf r}}\psi_n({\bf r})d^2{\bf r}\\
= \frac{1}{2\pi}\int_0^{\infty}\frac{\bar{r}^{n+1}e^{-\bar{r}^2/4}}{\sqrt{2\pi 2^n n!}}I(\bar{r})d\bar{r},
\label{FT}
\end{multline}
where $I(\bar{r}) = \int_0^{2\pi} e^{-ikl_{\mathrm{B}}\bar{r}\cos(\xi-\phi)}e^{in\phi}d\phi$, $\xi$ being the angle of ${\bf k}$ and $\bar{r} \equiv r/l_{\mathrm{B}}$. Making the variable change $\phi^{\prime} = \phi-\xi$ and defining $\beta \equiv -kl_{\mathrm{B}}\bar{r}$ we write
\begin{multline}
I(\bar{r}) = e^{i n\xi}\int_{-\xi}^{2\pi-\xi} e^{i\beta\cos\phi^{\prime}}e^{in\phi^{\prime}}d\phi^{\prime}\\
=  e^{i n\xi}\bigg[\int_{0}^{2\pi} e^{i\beta\cos\phi^{\prime}}\cos(n\phi^{\prime}) d\phi^{\prime}\\+i\int_{0}^{2\pi} e^{i\beta\cos\phi^{\prime}}\sin (n\phi^{\prime}) d\phi^{\prime}\bigg]\\
= 2e^{i n\xi} \int_{0}^{\pi} e^{i\beta\cos\phi^{\prime}}\cos(n\phi^{\prime}) d\phi^{\prime} = 2\pi e^{i n\xi}i^n J_n(\beta),
\label{Ir}
\end{multline}
where to obtain the second equality we used $e^{in\phi^{\prime}} = \cos(n\phi^{\prime})+i\sin(n\phi^{\prime})$ and changed the integration limits using the $2\pi$-periodicity of the integrand. The second integral containing $\sin(n\phi^{\prime})$ vanishes as the integrand is an odd function with respect to $\phi^{\prime} = \pi$. The third equality is due to the evenness of the integrand with respect to $\phi^{\prime} = \pi$ and the result of the final integral can be found in \cite{Integral table}, $J_n$ being the Bessel function of the first kind. Inserting Eq.~(\ref{Ir}) into Eq.~(\ref{FT}) we find
\begin{multline}
\tilde{\psi}_n({\bf k}) = \frac{l_{\mathrm{B}} i^n e^{i n\xi}}{\sqrt{2\pi 2^n n!}}\int_0^{\infty}\bar{r}^{n+1}e^{-\bar{r}^2/4}J_n(-kl_{\mathrm{B}}\bar{r})d\bar{r}\\
= l_{\mathrm{B}}(-i)^n \sqrt{\frac{2^{n+1}}{\pi n!}}  (kl_{\mathrm{B}})^n e^{in\xi} e^{-(kl_{\mathrm{B}})^2},
\label{psi_nk}
\end{multline}
where the result of the integral can again be found in \cite{Integral table}. 

Note that $\tilde{\psi}_n({\bf k})$ can be obtained from $\psi_n(\zeta)$ by changing $\zeta l_{\mathrm{B}} = (x + iy)$ to $-2il_{\mathrm{B}}^2 k e^{i\xi}$ and multiplying the result by $2l^2_B$ \cite{LLL expansion SM}. This formal analogy of the LLL wavefunctions in real and ${\bf k}$ spaces is related to the fact that the LLL states are eigenstates of a harmonic oscillator Hamiltonian of characteristic length $\sqrt{2}l_{\mathrm{B}}$~\cite{CCT} and that the Fourier transform operator corresponds to a quarter-period-long temporal evolution under a harmonic oscillator Hamiltonian~\cite{Andrews}.
 
\subsection{Mean square radius}

Focusing on a strictly two-dimensional case, the real-space density of atoms $\langle n({\bf r})\rangle_{{\rm tof}}$ measured through absorption imaging after the trapping potential and synthetic fields are suddenly turned off can be related to the momentum-space density $\langle \tilde{n}({\bf k})\rangle_{{\rm trap}}$ of trapped atoms before the sudden release as~\cite{cold atoms SM}
\be
\langle n({\bf r})\rangle_{{\rm tof}} = \langle \tilde{n}({\bf k})\rangle_{{\rm trap}} (d^2{\bf k}/d^2{\bf r})\simeq (M/\hbar t)^2\langle \tilde{n}({\bf k})\rangle_{{\rm trap}},
\label{tof equation}
\ee
where the ratio $(d^2{\bf k}/d^2{\bf r})$ between the infinitesimal area elements is approximated by the quantity $(M/\hbar t)^2$, which follows from the relation $\hbar{\bf k}\simeq M{\bf r}/t$ assuming a ballistic (free) expansion of the cloud for a time $t$ after all the fields are turned off. As mentioned in the main text, for the ballistic expansion condition to be valid, the interactions between particles should be negligible during the expansion, although the initial state can be a highly correlated one. This omission of interaction effects can be justified, for instance, whenever the system can be described within the LLL approximation as in the present case \cite{LLL expansion SM}.

The time-of-flight mean square radius $\langle r^2\rangle_{{\rm tof}} = \int \langle n({\bf r})\rangle_{{\rm tof}}r^2d^2{\bf r}/\int \langle n({\bf r})\rangle_{{\rm tof}}d^2{\bf r}$ can be calculated using $\hbar{\bf k}\simeq M{\bf r}/t$, $\langle n({\bf r})\rangle_{{\rm tof}} \simeq (M/\hbar t)^2\langle \tilde{n}({\bf k})\rangle_{{\rm trap}}$, and $\tilde{n}({\bf k}) = \tilde{\Psi}^{\dagger}({\bf k})\tilde{\Psi}({\bf k})$ \cite{cold atoms SM} as
\begin{multline}
\langle r^2\rangle_{{\rm tof}} \simeq \frac{1}{N}\bigg(\frac{\hbar t}{M}\bigg)^2\int \langle\tilde{\Psi}^{\dagger}({\bf k})\tilde{\Psi}({\bf k})\rangle_{\rm trap} k^2d^2{\bf k} \\
=\frac{1}{N}\bigg(\frac{\hbar t}{M}\bigg)^2\int \sum_{n,n^{\prime}} \tilde{\psi}_{n^{\prime}}^{\ast}({\bf k})\tilde{\psi}_n({\bf k})\langle a_{n^{\prime}}^{\dagger}a_n\rangle_{{\rm trap}}k^2d^2{\bf k} \\
=\frac{2\pi}{N}\bigg(\frac{\hbar t}{M}\bigg)^2\int \sum_{n} |\tilde{\psi}_n({\bf k})|^2\langle a_{n}^{\dagger}a_n\rangle_{{\rm trap}}k^2 kdk \\
= \frac{1}{2 N l^2_B}\bigg(\frac{\hbar t}{M}\bigg)^2\sum_n\frac{1}{n!}\langle a_{n}^{\dagger}a_n\rangle_{{\rm trap}}\int_0^{\infty}u^{n+1}e^{-u}du
\\
= \bigg(\frac{\hbar t}{\sqrt{2}Ml_{\mathrm{B}}} \bigg)^2 \frac{1}{N}\bigg(\frac{\langle L_z\rangle_{\rm trap}}{\hbar} + N\bigg),
\label{r2tof Lz SM}
\end{multline}
where the second line is obtained by expanding the Fourier transform $\tilde{\Psi}({\bf k}) = \int e^{-i{\bf k}\cdot{\bf r}}\Psi({\bf r})d^2{\bf r}/2\pi$ of the field operator $\Psi({\bf r})=\sum_n\psi_n({\bf r})a_n$, with $a_n$ ($a^{\dagger}_n$) destroying (creating) a particle in the single-particle state labelled by $n$ and $\tilde{\psi}_n({\bf k})$ is given by Eq.~(\ref{psi_nk}). The third line is obtained by performing the angular integration, which yields $2\pi\delta_{nn^{\prime}}$. We get the fourth line by making the variable change $u = 2(kl_{\mathrm{B}})^2$. The result of the integral is $(n+1)!$ and we finally obtain the last line by using the relations $\langle L_z\rangle/\hbar = \sum_n n\langle a_{n}^{\dagger}a_n\rangle$ and $N = \sum_n\langle a_{n}^{\dagger}a_n\rangle$ \cite{Ho Mueller SM}. Through a similar calculation the in-trap mean square radius can be found exactly to be
\be
\langle r^2\rangle_{{\rm trap}} = \frac{2l^2_B}{N}\bigg(\frac{\langle L_z\rangle_{\rm trap}}{\hbar} + N\bigg).
\ee
We see that the expansion of the cloud is self-similar since $\langle r^2\rangle_{{\rm tof}} \propto t^2\langle r^2\rangle_{{\rm trap}}$, which can be traced back to the same functional form of $\psi_n({\bf r})$ and $\tilde{\psi}_n({\bf k})$ \cite{LLL expansion SM}.

\section{Numerical Methods}

\subsection{\label{subsec:MC}Monte Carlo algorithm}
Results presented in Figs.~(1) and (2) in the main text required the computation of $\braket{r^2}$ for different quasihole states. For this purpose, we take advantage of the so-called Laughlin plasma analogy, in which the absolute square of the manybody wave function $|\Psi|^{2}$ is interpreted as the Boltzmann factor $e^{-\beta U}$ of a classical system of 2D charged particles at the fictitious temperature $\beta = 2 \nu$~\cite{Laughlin paper SM}. This allows us to recast the expectation value of several observables --notably those which only depend on the particle coordinates-- as an average over the probability distribution $e^{-\beta U}$. In particular, the mean square radius reads
\begin{equation}
\begin{split}
& \bra{\Psi_{\mathrm{qh}}(\{\zeta_{i}\},\{\mathcal{R}_{j}\})} \frac{1}{N} \sum^{N}_{i=1} r^{2}_{i} \ket{\Psi_{\mathrm{qh}}(\{\zeta_{i}\},\{\mathcal{R}_{j}\})} \\
&= \frac{\int d \zeta_{1} d \zeta_{1}^{\ast}\dots d \zeta_{N} d \zeta_{N}^{\ast}|\zeta_{1}|^{2} e^{-\beta U (\{\zeta_{i}\},\{\mathcal{R}_{j}\})} }{\int d \zeta_{1} d \zeta_{1}^{\ast}\dots d \zeta_{N} d \zeta_{N}^{\ast} e^{-\beta U (\{\zeta_{i}\},\{\mathcal{R}_{j}\})} }.
\end{split}
\label{eq:mean_squared_radius}
\end{equation}

We compute the integrals in Eq.~(\ref{eq:mean_squared_radius}) through the Metropolis Monte Carlo (MC) approach~\cite{Metropolis,Krauth_book}, which is an established method to study several properties of fractional quantum Hall (FQH) systems, including density profiles and pair correlation functions~\cite{PRB33_2221 SM,Kjonsberg}. At each MC step, all particles are moved by random displacements, and the move is accepted if $e^{-\beta\Delta U}$ is larger than a random number between zero and one, where $\Delta U$ is the energy change due to the move. The observables determined through this method are exact within statistical errors.

\subsection{Exact diagonalization}
\begin{figure}[b]
\includegraphics[width=0.43\textwidth]{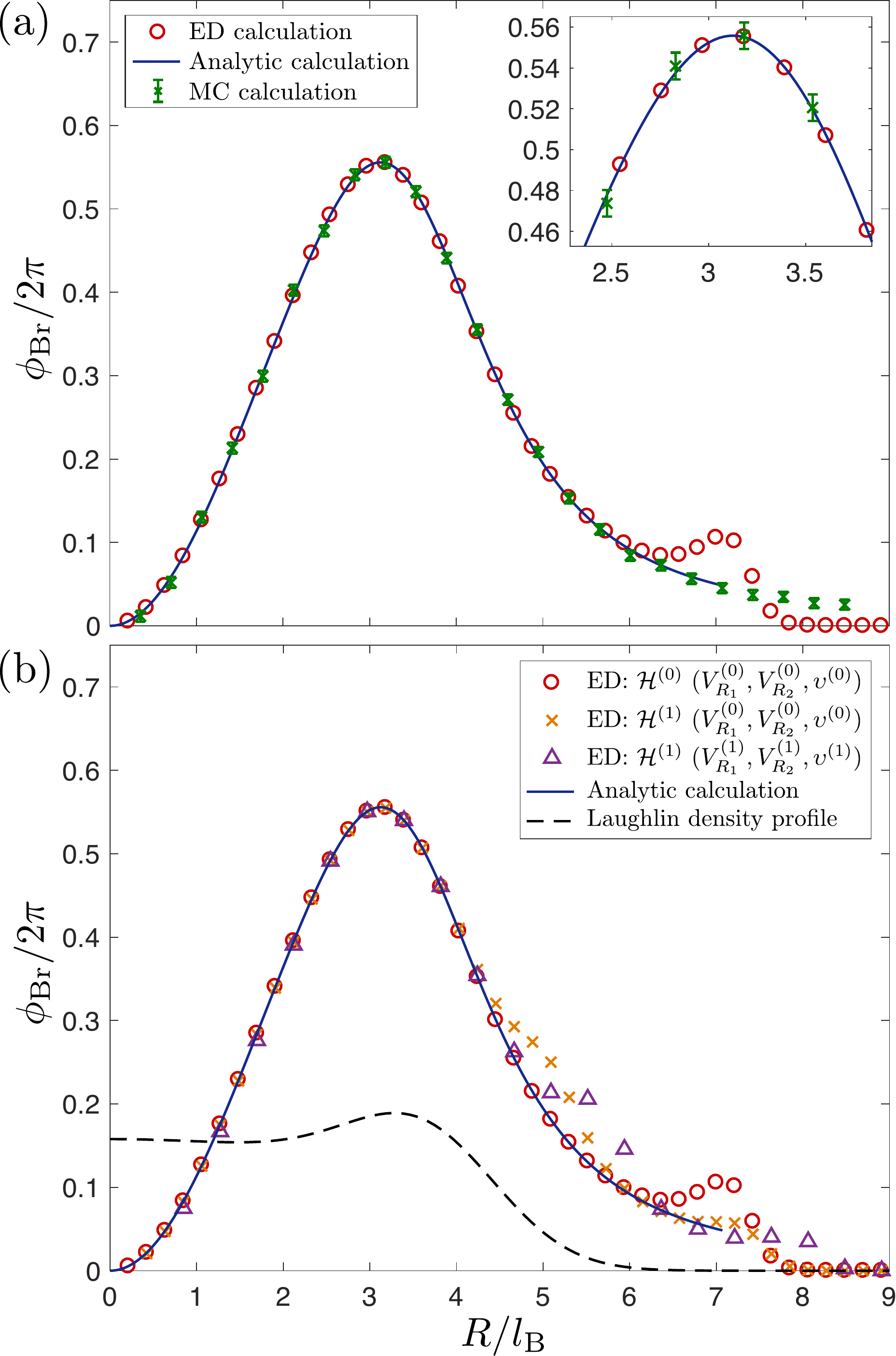}
\caption{(a) Comparison between the braiding phase $\phi_{\rm Br}$ as a function of $R$ obtained via ED method (red circles), analytic calculations for the ansatz wave functions (blue solid line) and MC method (green crosses) for a system of $N=6$ particles at filling $\nu = 1/2$ (one quasihole is taken to be in the center for the two-quasihole state). Parameters used in the diagonalization are $V_{R_{1}} = V_{R_{2}} = V_{\mathrm{int}}$ and $\upsilon = 10^{-4} V_{\mathrm{int}}$. The inset shows a zoom of a part of the curve to display the error bars of MC results. (b) Study of the dependence of the ED results for the braiding phase on both the Hilbert space size and the Hamiltonian parameters. Results obtained with parameters $(V^{(0)}_{R_{1}}=V^{(0)}_{R_{2}} = V_{\mathrm{int}}, \upsilon^{(0)} = 10^{-4} V_{\mathrm{int}})$ on the ``Jack'' Hilbert space $\mathcal{H}^{(0)}$ [those in (a)] are given by red circles and on the larger Hilbert space $\mathcal{H}^{(1)} \supset \mathcal{H}^{(0)}$ by orange crosses. Results obtained with stronger potential parameters $(V^{(1)}_{R_{1}}=V^{(1)}_{R_{2}} = 100 V_{\mathrm{int}}, \upsilon^{(1)} = 5 \times 10^{-4} V_{\mathrm{int}})$ on $\mathcal{H}^{(1)}$ are given by purple triangles. Note that in this last case, the deviations with respect to the analytical calculation shift towards larger $R$. Radial density profile of the $\nu=1/2$ Laughlin state for $N=6$ particles is shown in arbitrary units by the dashed line.}
\label{fig:EDvsMC_and_Hcutoff}
\end{figure}

To further support the results obtained by means of the above presented MC technique, we perform exact diagonalization (ED) calculations in the LLL approximation [see Fig. \ref{fig:EDvsMC_and_Hcutoff}(a)].
More precisely, we diagonalize the Hamiltonian $H_{\rm qh} = H_{\mathrm{FQH}} + V_{\mathrm{qh}} + V_{\mathrm{trap}}$ in the LLL, where $V_{\mathrm{qh}} = \sum_i[V_{R_{1}} \delta (\mathbf{r}_{i} - {\bf R}_{1}) + V_{R_{2}} \delta (\mathbf{r}_{i} - {\bf R}_{2})]$ describes the repulsive potentials associated with the two laser beams used to pin the quasiholes and the weak trap potential is chosen to be a harmonic one, which can be given in the LLL as $V_{\mathrm{trap}} = \upsilon L_{z}$ up to a constant energy shift. Typical values of the Hamiltonian parameters are $\upsilon = 1$--$5 \times 10^{-4} V_{\mathrm{int}}$ and $V_{R_{1}} = V_{R_{2}} = 1$--$100 V_{\mathrm{int}}$, in which $V_{\mathrm{int}} \equiv g_{\mathrm{int}}/ 2 \pi l^{2}_{\mathrm{B}}$ represents the characteristic energy scale of the system.

Regarding the Hilbert spaces over which we diagonalize the Hamiltonians, they are constructed by taking advantage of the possibility to decompose quasi-hole states in terms of Jack polynomials~\cite{PRL100_246802}. This choice is found to give extremely accurate results as long as both lasers are located reasonably inside the FQH cloud [see Fig. \ref{fig:EDvsMC_and_Hcutoff}(b)]. On the other hand, by enlarging the Hilbert space, small variations in the ground states are observed when one of the two lasers starts acting on the tail of the FQH cloud instead of on its bulk. 

From the physical point of view, these deviations of the numerical ground states with respect to the ansatz quasihole wave functions can be associated with the edge excitations that are created in addition to the quasihole when the pinning potential is located close to the edge of the cloud. Within the perspective of this work, however, such deviations are not relevant for two reasons. First, our proposal correctly applies only when the two quasiholes are both located in the bulk of the cloud, which is the region where the braiding and statistical phases have most of their meaning. Second, the range of $R = |{\bf R}_{1}-{\bf R}_{2}|$ values for which these deviations appear can be reduced if higher values of the laser strength --together with stronger harmonic confinements-- are considered.

As a final step, we test the possibility of pinning quasiholes with more realistic potentials by modeling the laser beams as finite-width Gaussians. Overlaps larger than $95\%$ between the numerical ground state and the required quasihole analytic wavefunction are observed for a relevant range of parameters. Moreover, the use of hard-wall confinements is found to be helpful in selecting the number of quasiholes pinned by each laser --see Ref.~\cite{EMIC_ring_fqh}, where repulsive square-well potentials centered in the origin are considered. For instance, the ground state of a $\nu = 1/2$ system experiencing a hard-wall potential~\cite{EMIC_disk_fqh} of parameters $V_{ext} = 100 V_{\mathrm{int}}$ and $R_{ext} = 5.2 \sqrt{2} l_{\mathrm{B}}$ and a repulsive Gaussian potential of the form
\begin{equation}
V_{G}(r) = (V_{R_{1}}/\sqrt{2 \pi} \sigma) \mathrm{exp} [- (r-R_{1})^{2}/2\sigma^{2}] ,
\end{equation}
with $R_{1}=0$ and $V_{R_{1}} = V_{\mathrm{int}}$, has an overlap with $\Psi_{\rm 1qh}(R_1 = 0)$ larger than $98\%$ when $\sigma \leq 0.5 l_{\mathrm{B}}$.

\section{More general quasihole configurations}
In the main text we considered pinning potentials localizing quasiholes in peculiar configurations with high symmetry. We now relax these assumptions in two directions: first by treating more general configurations, and second by letting quasiholes fluctuate around their reference positions. These points may be relevant in view of experimental realizations, where uncertainties in the quasihole localization would naturally be present.

\begin{figure}[b]
\includegraphics[width=0.47\textwidth]{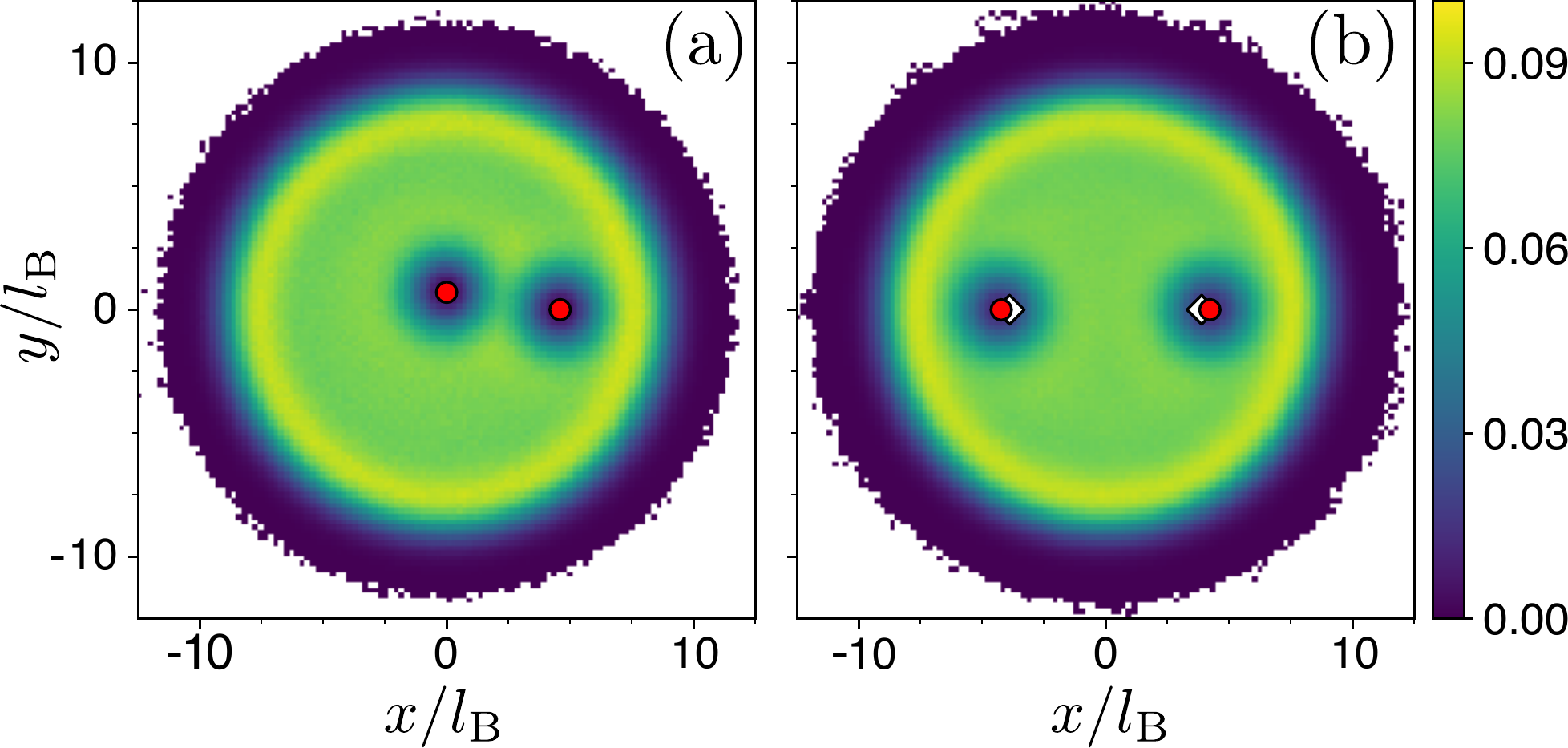}
\caption{(a) Density profile for the state with two quasiholes pinned at $\mathbf{R}_{1} = (3.25 \sqrt{2} l_{\rm{B}} , 0)$ and $\mathbf{R}_{2} = (0, 0.5 \sqrt{2} l_{\rm{B}})$. (b) Density profile obtained through the modified sampling for quasiholes fluctuating around the reference positions $\mathbf{R}_{1} = - \mathbf{R}_{2} =(2.75 \sqrt{2} l_{\mathrm{B}},0)$ (white diamonds). The average quasihole positions (red circles) are shifted outwards with respect to the reference ones, due to interaction with the classical particles forming the plasma.}
\label{fig:generic_qh_configs}
\end{figure}

For two quasiholes pinned at generic positions $\mathbf{R}_{1}$ and $\mathbf{R}_{2}$, the exchange process can be represented as a $\pi$--rotation plus a topologically irrelevant translation by $\mathbf{R}_{1}+\mathbf{R}_{2}$. The expression for the statistical phase, Eq.~$(8)$ in the main text, is then generalized as
\begin{equation}
\begin{split}
\phi_{\rm st}(\mathbf{R}_{1},\mathbf{R}_{2}) =& \frac{\pi}{\hbar} \bigg( \langle L_z\rangle^{\rm 2qh}-\langle L_z\rangle^{\rm 1qh}_{(1)} -\langle L_z\rangle^{\rm 1qh}_{(2)} \bigg) \\
& + \pi m \frac{N(N-1)}{2} ,
\end{split}
\end{equation}
where $\langle L_z\rangle^{\rm 1qh}_{(i)}$ is the average angular momentum for the state containing only the quasihole in $\mathbf{R}_{i}$. Therefore, for configurations of this kind, one needs to evaluate three different average values of $L_{z}$. We compute $\phi_{\rm{st}}$ through the MC algorithm for $\mathbf{R}_{1} = (3.25 \sqrt{2} l_{\rm{B}} , 0)$ and $\mathbf{R}_{2} = (0, 0.5 \sqrt{2} l_{\rm{B}})$ [see Fig.~\ref{fig:generic_qh_configs} (a)]. The result, $\phi_{\rm{st}} = (0.499 \pm 0.003) \pi/\hbar$, is fully compatible with $\pi / (2 \hbar)$, which is the expected value when $\mathbf{R}_{1}$ and $\mathbf{R}_{2}$ coordinates are sufficiently far from each other and from the cloud boundary.

Due to the finite size of the depletion regions characterizing quasiholes, thermal fluctuations may induce small deviations in their positions. We include this possibility in our MC algorithm by letting the quasiholes move during the sampling and we keep each of them close to its reference position by means of an additional harmonic trap in the plasma-analogy potential energy $U$. For each choice of the plasma parameters, we set the characteristic length scale of these traps such that fluctuations in the quasihole positions are of the order of a fraction of their extension (proportional to $l_{\mathrm{B}}$).
As a concrete example we evaluate the statistical phase characterizing quasiholes fluctuating around $\mathbf{R}_{1}= - \mathbf{R}_{2} = (2.75 \sqrt{2} l_{\mathrm{B}},0)$ with standard deviation $0.4 l_{\mathrm{B}}$ [see Fig.~\ref{fig:generic_qh_configs} (b)]. The result, $\phi_{\rm{st}} = (0.498 \pm 0.004) \pi/\hbar$, does not show deviations with respect to the case of fixed quasiholes positions treated in the main text.

